\documentclass[sigconf, nonacm]{acmart}
\usepackage{tablefootnote}
\AtBeginDocument{%
  }

\setcopyright{acmlicensed}
\copyrightyear{2025}
\acmYear{2025}
\acmDOI{XXXXXXX.XXXXXXX}
\acmConference[ICAIF ’25]{ACM International Conference on AI in Finance (ICAIF ’25)}{November 15--18, 2025}{Singapore}
\acmISBN{978-1-4503-XXXX-X/2018/06}

\begin{document}

\title{FinAgentBench: A Benchmark Dataset for Agentic Retrieval in Financial Question Answering}


\author{Chanyeol Choi}
\authornote{Corresponding Author.}
\email{jacobchoi@linqalpha.com}
\affiliation{%
  \institution{LinqAlpha}
  \country{United States}
}

\author{Jihoon Kwon}
\email{jihoonkwon@linqalpha.com}
\affiliation{%
  \institution{LinqAlpha}
  \country{United States}
}

\author{Alejandro Lopez-Lira}
\email{alejandro.lopez-lira@warrington.ufl.edu}
\affiliation{%
  \institution{University of Florida}
  \country{United States}
}

\author{Chaewoon Kim}
\email{kcw0826@snu.ac.kr}
\affiliation{%
  \institution{LinqAlpha}
  \country{United States}
}

\author{Minjae Kim}
\email{minjaekim@linqalpha.com}
\affiliation{%
  \institution{LinqAlpha}
  \country{United States}
}

\author{Juneha Hwang}
\email{junehahwang@linqalpha.com}
\affiliation{%
  \institution{LinqAlpha}
  \country{United States}
}

\author{Jaeseon Ha}
\email{jaeseonha@linqalpha.com}
\affiliation{%
  \institution{LinqAlpha}
  \country{United States}
}

\author{Hojun Choi}
\email{hojunchoi@linqalpha.com}
\affiliation{%
  \institution{LinqAlpha}
  \country{United States}
}

\author{Suyeol Yun}
\email{suyeolyun@linqalpha.com}
\affiliation{%
  \institution{LinqAlpha}
  \country{United States}
}

\author{Yongjin Kim}
\email{jinkim@linqalpha.com}
\affiliation{%
  \institution{LinqAlpha}
  \country{United States}
}

\author{Yongjae Lee}
\email{yongjaelee@unist.ac.kr}
\affiliation{%
  \institution{UNIST}
  \country{Republic of Korea}
}

\renewcommand{\shortauthors}{Choi et al.}

\begin{abstract}
Accurate information retrieval (IR) is critical in the financial domain, where investors must identify relevant information from large collections of documents.
Traditional IR methods—whether sparse or dense—often fall short in retrieval accuracy, as it requires not only capturing semantic similarity but also performing fine-grained reasoning over document structure and domain-specific knowledge.
Recent advances in large language models (LLMs) have opened up new opportunities for \emph{retrieval with multi-step reasoning}, where the model ranks passages through iterative reasoning about which information is most relevant to a given query.
However, there exists no benchmark to evaluate such capabilities in the financial domain.
To address this gap, we introduce \textsc{FinAgentBench}, the first large-scale benchmark for evaluating retrieval with multi-step reasoning in finance -- a setting we term \emph{agentic retrieval}.
The benchmark consists of 26K expert-annotated examples on S\&P-500 listed firms and assesses whether LLM agents can (1) identify the most relevant document type among candidates, and (2) pinpoint the key passage within the selected document.
Our evaluation framework explicitly separates these two reasoning steps to address context limitations.
This design enables to provide a quantitative basis for understanding retrieval-centric LLM behavior in finance.
We evaluate a suite of state-of-the-art models and further demonstrated how targeted fine-tuning can significantly improve agentic retrieval performance.
Our benchmark provides a foundation for studying retrieval-centric LLM behavior in complex, domain-specific tasks for finance.
\end{abstract}

\ccsdesc[500]{Information systems~Information retrieval}
\ccsdesc[500]{Information systems~Retrieval models and ranking}
\ccsdesc[300]{Information systems~Language models}

\keywords{Information Retrieval, Generative Retrieval, Large Language Models}


\maketitle

\section{Introduction}

Information Retrieval (IR) is a foundational research field that studies how to effectively search for and retrieve relevant information from large-scale text collections~\citep{salton1968automatic, schutze2008introduction}.
Its practical importance in real-world applications has made it one of the central and long-standing areas in computer science since the early days of computing~\citep{mitra2000information}.
IR has evolved from sparse, term-frequency–based methods~\citep{sparck1972statistical}, which rely on exact keyword matches, to dense neural retrieval models that embed text into continuous latent spaces to capture deeper semantic~\citep{karpukhin2020dense}.
In finance, accurate retrieval is critical, as investors depend on precise access to vast filings and reports to make high-stakes, time-sensitive decisions.
To support this need, finance-specific IR benchmarks have been developed~\citep{islam2023financebench, choi2025FinDER}, enabling rigorous evaluation of both sparse and dense retrievers in this complex, data-rich domain.

\begin{figure}[t]
    \centering
\includegraphics[width=0.49\textwidth]{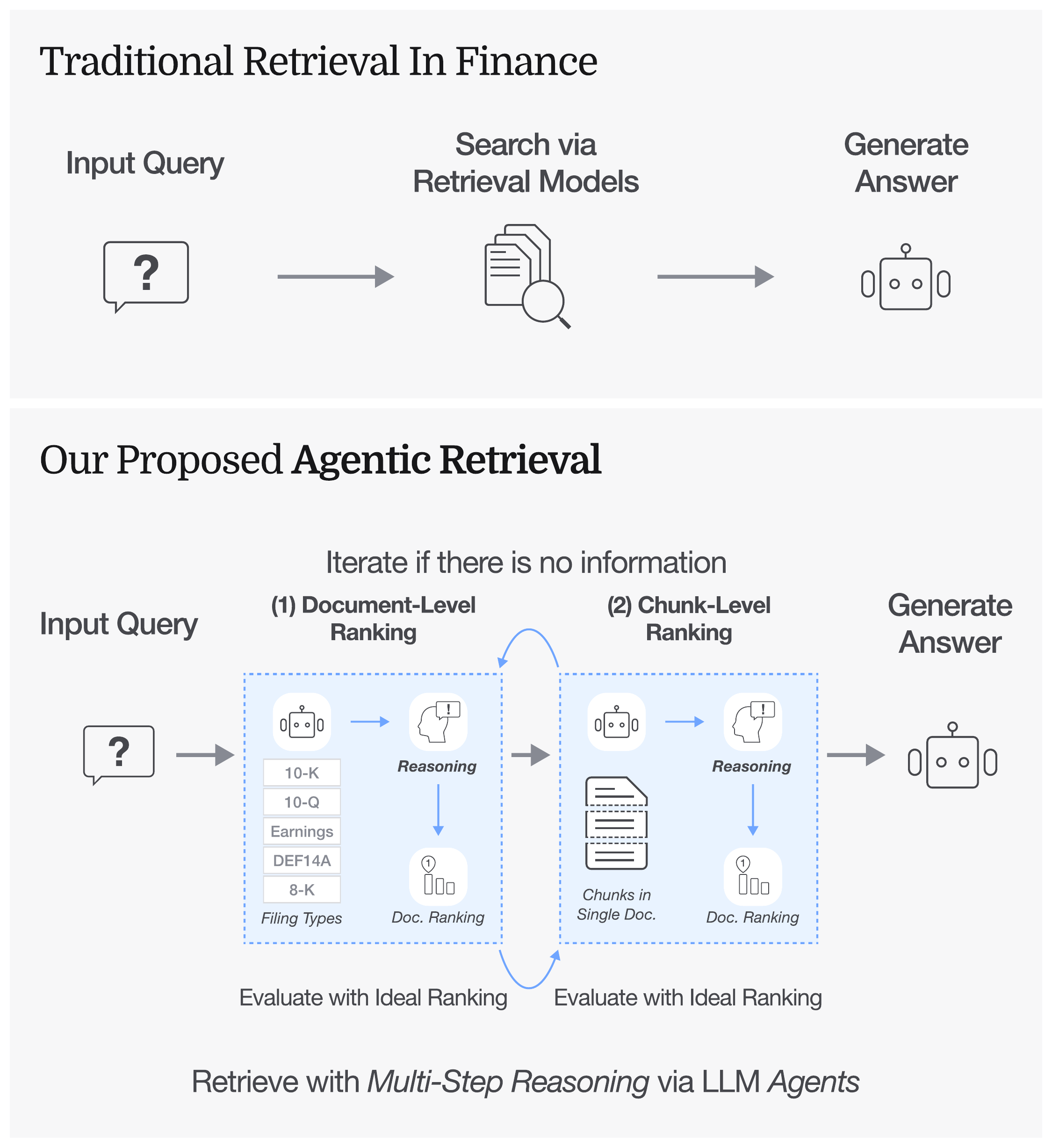}
\caption{
Comparison between traditional retrieval and our proposed \emph{agentic retrieval}. While traditional systems retrieve in a single step, \emph{agentic retrieval} decomposes the task into document and chunk ranking via multi-step reasoning.
}
\label{fig:teasor}
\end{figure}

However, recent studies reveal persistent accuracy ceilings for both sparse and dense retrieval methods~\citep{liu2024robust, chan2025don}, especially in domains that demand fine-grained understanding and structured reasoning over complex documents.
To overcome these limitations, recent research has turned to large language models (LLMs), which bring strong language understanding and the ability to process long contexts~\citep{gao2023retrieval, luo2024large, zhu2023large}.
One promising direction is \emph{generative retrieval}, where an LLM generates the index of the most relevant document given a query and document collection~\citep{nguyen2023generative}.
These results show impressive gains in performance, indicating a shift toward retrieval systems that embed deeper reasoning capabilities~\citep{cai2025exploring, pradeep2023does}.

In the financial domain, this shift is particularly relevant.
Financial documents are typically long and dense, often requiring multiple steps of reasoning: first to identify which document type (e.g., 10-K, earnings transcript) best matches the information need, and then to locate the specific evidence within it.
As such, simple one-shot retrieval is often insufficient—accurate performance hinges on multi-step reasoning that reflects how professionals actually search for information.
Despite the practical importance of such multi-step reasoning, no benchmark currently exists to evaluate LLMs in this setting. This leaves open critical questions about whether LLMs can serve as effective retrieval agents in high-stakes domains like finance—where precision, explainability, and structured navigation are essential.

To address this gap, we introduce the first large-scale benchmark, \textsc{FinAgentBench}, for evaluating generative retrieval systems in the financial domain, focusing on a setting we term \emph{agentic retrieval}.
Unlike prior benchmarks that assess retrieval in a single stage, our proposed benchmark evaluates the ability of LLM agents to reason and retrieve the relevant information through a two-stage pipeline: (1) identifying which document to retrieve, evaluated via document-level ranking, and (2) determining which part of the document to focus on, evaluated via chunk-level ranking.
The benchmark contains about 26K samples and is constructed from real-world financial documents paired with expert-written queries and annotations, reflecting authentic use cases faced by professional investors.
By capturing both retrieval accuracy and reasoning depth, it provides a foundation for systematically analyzing the strengths and limitations of LLM-based generative retrieval in high-stakes domains like finance.

The main contribution of this work is three-fold:
\begin{itemize}
  \item We propose \textsc{FinAgentBench}, to the best of our knowledge the first large-scale benchmark for evaluating agentic retrieval in finance, featuring about 26K expert-annotated samples across document- and chunk-level ranking tasks.
  \item We evaluate a range of state-of-the-art LLMs on our benchmark, revealing their performance in accuracy and reasoning when applied to real-world financial retrieval scenarios.
  \item We further investigate the impact of fine-tuning the LLMs on agentic retrieval tasks, demonstrating that targeted supervision can substantially improve both document selection and chunk-level reasoning performance.

\end{itemize}

\section{Related Work}

\paragraph{\textbf{Retrieval Systems.}}
IR systems has progressed from early sparse methods like TF–IDF and BM25~\citep{Salton1975,Robertson1994}, which rely on exact lexical overlap, to dense dual encoders that map queries and documents into a shared semantic space. Notable advances such as Dense Passage Retrieval (DPR)~\citep{Karpukhin2020DPR} and E5-Mistral~\citep{Wang2024E5} have improved performance on open-domain QA by capturing deeper semantics.
Despite these gains, both sparse and dense retrievers struggle with multi-hop reasoning and long-context queries—highlighting inherent limitations in fixed retrieval pipelines.

To address this, \emph{generative retrieval} reframes the task as sequence generation, enabling models to directly produce relevant document identifiers or content~\citep{li269303210matching}. Recent approaches such as CAG~\citep{CAG2024}, and Infinite Retrieval~\citep{Ye2025InfiniteRetrieval} leverage LLMs’ language modeling and attention capabilities by caching the documents to bypass traditional indexing.
In parallel, agentic reasoning frameworks like ReAct~\citep{Yao2023ReAct} incorporate tool use and step-by-step reasoning, suggesting the potential of the LLMs to iteratively decide what to retrieve and why—bridging retrieval with planning and decision-making.

\paragraph{\textbf{Finance-Retrieval Benchmarks.}}
Retrieval in the financial domain presents unique challenges: documents are long, queries are complex, and precision is critical. Existing datasets such as FinQA~\citep{Chen2021FinQA}, TAT-QA~\citep{Zhu2021TATQA}, and FiQA~\citep{Maia2018FiQA} address specific financial reasoning needs, while more recent corpora like FinanceBench~\citep{islam2023financebench} and FinDER~\citep{choi2025FinDER} support open-domain and retrieval-augmented generation tasks.
However, these benchmarks still assume retrieval as a fixed subroutine, typically relying on vector search or keyword search—and do not evaluate whether LLMs can reason about what to retrieve or where to look within long documents.

Our proposed \textsc{FinAgentBench} fills this gap by directly evaluating agentic retrieval in finance. It assesses whether an LLM agent can (1) generate the correct document identifier and (2) rank the most relevant chunks within that document, providing a rigorous testbed for end-to-end reasoning-driven retrieval in high-stakes settings.

\begin{figure}[t]
    \centering
\includegraphics[width=0.49\textwidth]{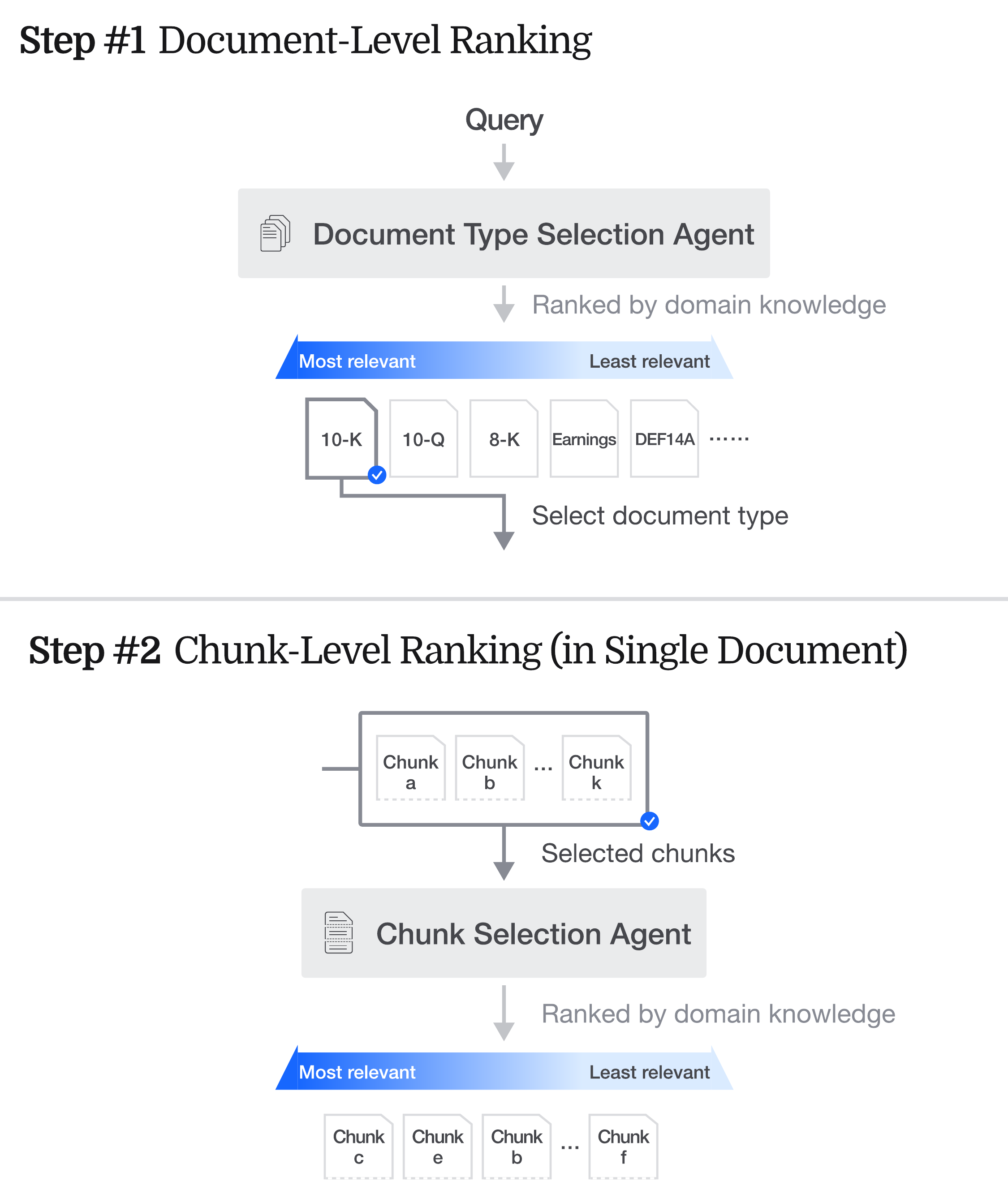}
    \caption{
Agentic retrieval pipeline in \textsc{FinAgentBench}. The process consists of two stages. 
(\textit{Top}) Given a natural-language query, the \textbf{Document Type Selection Agent} ranks five SEC filing types (\textsc{10-K}, \textsc{10-Q}, \textsc{8-K}, earnings transcripts, and DEF-14A) and selects the most relevant type.
(\textit{Bottom}) The corresponding document is segmented into paragraph-level chunks, from which the \textbf{Chunk Selection Agent} identifies the top-$k$ passages.
}
\label{fig:pipeline}
\end{figure}

\section{\textsc{FinAgentBench} Dataset}

In this section, we describe the construction of \textsc{FinAgentBench}, a large-scale benchmark for evaluating agentic retrieval in finance. Section~\ref{ssec:problem-setup} outlines motivates the need for agentic retrieval, followed by the task formulation and reasoning pipeline.
Section~\ref{ssec:data-collection} details the document and query collection process. Section~\ref{ssec:data-annotation} describes the annotation protocol and dataset statistics.

\subsection{Problem Setup}
\label{ssec:problem-setup}
To retrieve accurate information, financial retrieval requires multi-step reasoning due to both the volume of data and the regularity of financial disclosures~\citep{choe2025hierarchical}.
When leveraging the reasoning capabilities of large language models to improve accuracy in retrieval, the extensive length and redundancy of financial documents—where even a single 10-K can exceed hundreds of pages—pose a significant challenge, making it inefficient to process all content without any filtering or prioritization~\citep{reddy2024docfinqa}.
The computational cost and latency associated with brute-force reasoning over all available text make such an approach impractical, especially when dealing with many documents across multiple companies and periods.

Therefore, to maintain efficiency, a system should first select the document type most likely to contain the answer—feasible information because filings follow predictable conventions, with different information consistently organized by document type (e.g., risk factors in 10-Ks, strategic commentary in earnings calls).
It should then identify the relevant chunk or passage within the selected document. This motivates a two-stage retrieval process—document selection followed by passage selection—which we term \emph{agentic retrieval}, as it reflects the sequential reasoning steps taken by experts.

Figure~\ref{fig:pipeline} provides an overview of the \emph{agentic retrieval} workflow we benchmark.
At test time, the system is provided with a natural-language query $q$, typically issued by a professional investor about a specific firm, along with a document collection $\mathcal{D}$ containing over 35K U.S.\ corporate disclosures spanning from 2010 to 2024.
In \textsc{FinAgentBench}, we include 10-K, 10-Q, 8-K, earnings call transcripts, and DEF 14A proxy statements, some of the most commonly used public filings that are crucial for financial retrieval.
The retrieval task follows a two-stage reasoning pipeline:

\paragraph{\textbf{Stage 1: Document-Level Ranking.}}
Rather than searching the entire corpus, the agent first identifies the document type most likely to contain the answer. Given the type set 
\begin{equation}
\mathcal{T} = \{\text{10-K}, \text{10-Q}, \text{8-K}, \text{Earnings}, \text{DEF14A}\}
\end{equation}
, the model produces a ranking over $\mathcal{T}$. This stage evaluates the model’s understanding of finance-specific reporting conventions—for instance, risk factors typically appear in 10-Ks, while shareholder proposals are found in DEF-14A filings.

\begin{figure*}[t]
    \centering
\includegraphics[width=1.0\textwidth]{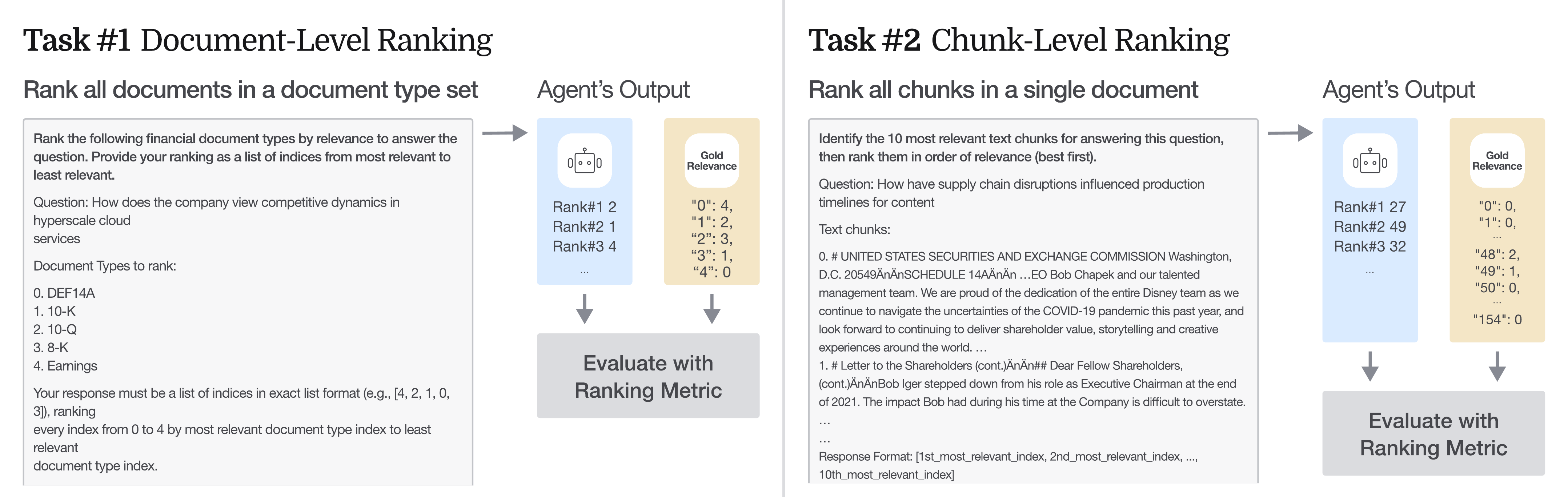}
\caption{
Examples of the two retrieval tasks in \textsc{FinAgentBench}.
The benchmark comprises: 
(\textit{Left}) Document-level ranking, where the agent ranks five SEC document types based on their relevance to the input query. All document types are ordered by relevance, and the full ranking is used as the gold label during evaluation.
(\textit{Right}) Chunk-level ranking, where the agent selects and orders the top-$k$ most relevant passages from the selected document. Each chunk is annotated with a ground-truth relevance label—0 (irrelevant), 1 (partially relevant), or 2 (directly relevant)—and predictions are evaluated using MRR, MAP, and nDCG.
}
\label{fig:tasks}
\end{figure*}

\paragraph{\textbf{Stage 2: Chunk-Level Ranking.}}
The selected document $(d_{t^\star} \in \mathcal{D})$ is split into non-overlapping passages $\mathcal{C}(d_{t^\star}) = \{c_1, \ldots, c_M\}$. The agent ranks these chunks and returns the top $k$ passages indexed as $\langle c_{(1)}, \ldots, c_{(k)} \rangle$.

\paragraph{\textbf{Annotations.}}
Each query in \textsc{FinAgentBench} is generated and annotated by domain experts. At the document-type level, a gold label $t^{\text{G}}$ in $\mathcal{T}$ is provided with an gold ranking .
At the chunk level, every chunk element $\mathcal{C}_{(i)}$ in $\mathcal{C}(d_{t^\star})$ is annotated as $\mathcal{C}_{(i)}^{\text{G}}$ with a gold relevance score.
For annotating those relevance scores, we followed TREC Eval~\citep{palotti2019}: 0 (irrelevant), 1 (partially relevant), and 2 (directly relevant).

\paragraph{\textbf{Evaluation.}}
We evaluate retrieval accuracy at both stages using standard top-\(k\) metrics: MRR, MAP, and nDCG. Success requires accurate performance at both the document-type and chunk levels.

\subsection{Data Collection}
\label{ssec:data-collection}

We construct \textsc{FinAgentBench} from real-world corporate filings and expert-authored queries to simulate authentic retrieval tasks in finance. The dataset comprises two main components: a large corpus of U.S.\ corporate documents and a set of fine-grained, expert-written queries categorized by information need.

\paragraph{\textbf{Document Collection}}
We collect filings from the U.S.\ Securities and Exchange Commission (SEC) EDGAR database, focusing on disclosures published between 2023 and 2024. Our selection includes approximately 3{,}000 publicly traded U.S.\ companies. For each firm, we retrieve five key document types frequently used by financial professionals: 10-K, 10-Q, 8-K, earnings call transcripts, and DEF-14A proxy statements. In total, the corpus consists of over 15{,}000 documents.

Each document is preprocessed into paragraph-level chunks. To preserve semantic and contextual integrity, we treat each table as a single unit and ensure the entire table—including surrounding textual context—is encapsulated within a single chunk. This strategy is particularly important in financial documents, where tabular data often conveys critical, high-density information.

In this paper, we curate a subset of the corpus by focusing exclusively on companies listed in the S\&P-500 index. This filtering yields a more targeted dataset comprising approximately 500 documents, which serves as the primary scope of analysis throughout the study.

\paragraph{\textbf{Query Collection}}

To simulate realistic information-seeking behavior, we employ two experienced domain experts to write queries grounded in 10 distinct categories: Analyst Q\&A, Management Commentary, Guidance, Industry \& Market, Investor Sentiment, Earnings Result \& Financials, Compensation, Macro \& Economics, Risks \& Challenges, and Operating Metrics.

For each company, both experts independently generate ten queries per category, resulting in 20 candidate queries. The final set of ten queries per firm is selected through cross-validation between the annotators to ensure coverage, diversity, and clarity. This process results in a collection of high-quality, expert-generated queries that span a broad range of investor concerns and reflect real-world analytic workflows.

\begin{table}[t]
  \centering
  \caption{Evaluation of reasoning LLMs on the \emph{Document Ranking} task.}
  \label{tab:doc-rank}
  \begin{tabular}{lccc}
    \toprule
    \textbf{Model} & \textbf{nDCG@5} & \textbf{MAP@5} & \textbf{MRR@5} \\
    \midrule
    GPT-o3            & 0.770 & 0.829 & 0.875 \\
    Claude-Opus-4     & 0.773 & 0.840 & 0.875 \\
    Claude-Sonnet-4   & \textbf{0.783} & \textbf{0.849} & \textbf{0.892} \\
    \bottomrule
  \end{tabular}
\end{table}
\begin{table}[t]
  \centering
  \caption{Evaluation of reasoning LLMs on the \emph{Chunk Ranking} task.}
  \label{tab:chunk-rank}
  \begin{tabular}{lccc}
    \toprule
    \textbf{Model} & \textbf{nDCG@5} & \textbf{MAP@5} & \textbf{MRR@5} \\
    \midrule
    GPT-o3            & 0.351 & 0.257 & 0.538 \\
    Claude-Opus-4     & 0.418 & \textbf{0.307} & \textbf{0.568} \\
    Claude-Sonnet-4   & \textbf{0.419} & 0.296 & 0.567 \\
    \bottomrule
  \end{tabular}
\end{table}

\subsection{Data Annotation}
\label{ssec:data-annotation}

\begin{table*}[t]
  \centering
  \caption{Impact of reinforcement fine-tuning on \texttt{GPT-o4-mini} across both retrieval tasks.}
  \resizebox{0.8\textwidth}{!}{
  \label{tab:fine_tune}
  \begin{tabular}{lcccccc}
    \toprule
    & \multicolumn{3}{c}{\textbf{w/o fine-tuning}} 
    & \multicolumn{3}{c}{\textbf{w/ fine-tuning}} \\
    \cmidrule(lr){2-4}\cmidrule(lr){5-7}
    \textbf{Task} & nDCG@5 & MAP@5 & MRR@5 & nDCG@5 & MAP@5 & MRR@5 \\
    \midrule
    Document Ranking & 0.758 & 0.826 & 0.872 & \textbf{0.808} & \textbf{0.865} & \textbf{0.933} \\
    Chunk Ranking    & 0.345 & 0.256 & 0.526 & \textbf{0.371} & \textbf{0.274} & \textbf{0.587} \\
    \bottomrule
  \end{tabular}
  }
\end{table*}

To convert raw filings and expert-written queries into ranking-annotated data, we implemented a structured two-stage annotation pipeline aligned with the retrieval stages described in Section~\ref{ssec:problem-setup}. Figure~\ref{fig:tasks} illustrates example instances and the corresponding annotation procedure for each task.

In the first stage, document-type ranking, annotators are given a query and a fixed set of five SEC document types (\textsc{10-K}, \textsc{10-Q}, \textsc{8-K}, earnings transcripts, and DEF-14A). They are asked to rank the document types in order of relevance by rearranging them via a drag-and-drop interface.
In the second stage, annotators examine all paragraph-level chunks from the most relevant to the least relevant filing of the top-ranked document type and assign a graded relevance score (0, 1, or 2) to each passage based on how well it answers the query.

All annotations were conducted by finance professionals with relevant industry experience.
Each query was independently labeled by two annotators, and disagreements were resolved through a brief adjudication process.
This dual-annotator setup with adjudication effectively functions as a cross-validation mechanism, enhancing the reliability and consistency of the labels.

\section{Experiments}

We evaluate a range of reasoning-capable LLMs on \textsc{FinAgentBench} to assess their ability to perform agentic retrieval in finance. Our experiments focus on two subtasks: document-type ranking and chunk-level passage selection. We also study the effect of domain-specific fine-tuning on retrieval performance.

\subsection{Experimental Setup}
We benchmark three commercial LLMs—\texttt{GPT-o3}, \texttt{Claude-Opus-4}, and \texttt{Claude-Sonnet-4}—using zero-shot prompting. For each query, models are prompted to complete two tasks: (1) rank document types from a set of five SEC filing categories, and (2) rank paragraph-level chunks from the selected document.
We split both the document-type ranking and chunk-level relevance tasks into training and evaluation sets using an 80/20 split.

To assess the impact of domain adaptation, we additionally evaluate \texttt{GPT-o4-mini} before and after reinforcement fine-tuning provided by OpenAI~\footnote{\href{https://platform.openai.com/docs/guides/reinforcement-fine-tuning}{OpenAI Reinforcement Fine-Tuning}} on held-out randomly sampled 10\% of training splits from \textsc{FinAgentBench}.

Performance is measured using standard ranking metrics measure by top-\(5\) results: normalized Discounted Cumulative Gain (nDCG), Mean Average Precision (MAP), and Mean Reciprocal Rank (MRR). For chunk-level evaluation, we report metrics over the top ranked passages, evaluated against expert-annotated graded relevance scores.
All models operate in a retrieval-only setting without access to external tools or retrieval augmentation.

\subsection{Results}

\paragraph{\textbf{Document-Level Ranking.}}
Table~\ref{tab:doc-rank} reports the performance of various reasoning-capable LLMs on the document-type ranking task. All models are prompted to rank the five candidate SEC document types given a query. Among the models, \texttt{Claude-Sonnet-4} achieves the highest performance across all three metrics, with an nDCG of 0.783, MAP of 0.849, and MRR of 0.892. Notably, all three models perform well above random, suggesting that general-purpose LLMs have a strong prior over financial reporting structure. The margin between models, however, indicates that modeling subtle type-level cues benefits from architectural or scale differences.

\paragraph{\textbf{Chunk-Level Ranking.}}
Table~\ref{tab:chunk-rank} presents results for the chunk-level passage retrieval task, where the model is asked to identify the top-5 rank of the most relevant content within a selected document. Performance is generally lower than in the document-type setting, reflecting the increased complexity of fine-grained retrieval. \texttt{Claude-Sonnet-4} and \texttt{Claude-Opus-4} perform comparably, with Sonnet slightly ahead in nDCG (0.419) and Opus leading in MAP (0.307) and MRR (0.568). This suggests that chunk-level reasoning remains challenging, even for strong LLMs, likely due to the need for knowledge-intensive reasoning over the lengthy context of financial disclosures.

\paragraph{\textbf{Impact of Fine-Tuning.}}
Table~\ref{tab:fine_tune} evaluates the impact of reinforcement fine-tuning on \texttt{GPT-o4-mini} across both retrieval stages. Fine-tuning yields substantial gains: for document-type ranking, nDCG improves from 0.758 to 0.808, and MRR increases from 0.872 to 0.933. Similar improvements are observed in chunk ranking, with MRR rising from 0.526 to 0.587. These results demonstrate that domain-specific supervision significantly enhances retrieval accuracy, both in selecting the correct document type and in pinpointing relevant information within the document. This highlights the importance of adapting LLMs to financial reasoning tasks through task-aligned training signals.

\section{Conclusion}

We introduced \textsc{FinAgentBench}, a large-scale benchmark designed to evaluate agentic retrieval capabilities of large language models in the high-stakes domain of finance. The benchmark simulates realistic investor queries over a diverse set of corporate filings, requiring models to reason over both document types and intra-document content. 

Our experiments show that state-of-the-art LLMs already exhibit strong priors over financial reporting structure, achieving high performance on document-type ranking. However, their ability to identify relevant content at the chunk level remains limited, reflecting the inherent challenges of fine-grained reasoning in long and information-dense documents.

We further demonstrate that reinforcement-style fine-tuning with domain-specific supervision can substantially improve performance on both retrieval stages, highlighting the value of aligning LLMs with domain experts' annotations.

\textsc{FinAgentBench} provides a new foundation for studying end-to-end retrieval behaviors in complex domains. Future work may explore enhancing agentic retrieval performance and joint modeling of retrieval and generation for investment decision support.


\bibliographystyle{ACM-Reference-Format}
\bibliography{sample-base}


\end{document}